\documentclass[twocolumn,aps,showpacs,prl,superscriptaddress]{revtex4}
\usepackage{graphicx}
\usepackage{amsmath}
\begin{document}
\title{Stationary states and energy cascades in inelastic gases}
\author{E.~Ben-Naim}\email{ebn@lanl.gov}
\affiliation{Theoretical Division and Center for Nonlinear Studies,
Los Alamos National Laboratory, Los Alamos, New Mexico 87545}
\author{J.~Machta}\email{machta@physics.umass.edu}
\affiliation{Department of Physics, University of Massachusetts,
Amherst, Massachusetts 01003}
\begin{abstract}

We find a general class of nontrivial stationary states in inelastic
gases where, due to dissipation, energy is transfered from large
velocity scales to small velocity scales. These steady-states exist
for arbitrary collision rules and arbitrary dimension. Their signature
is a stationary velocity distribution $f(v)$ with an algebraic
high-energy tail, $f(v)\sim v^{-\sigma}$. The exponent $\sigma$ is
obtained analytically and it varies continuously with the spatial
dimension, the homogeneity index characterizing the collision rate,
and the restitution coefficient.  We observe these stationary states
in numerical simulations in which energy is injected into the system
by infrequently boosting particles to high velocities. We propose that
these states may be realized experimentally in driven granular
systems.

\end{abstract}
\maketitle

Energy dissipation has profound consequences in granular media. It is
responsible for collective phenomena such as hydrodynamic
instabilities \cite{gz,km}, shocks \cite{rbss,slk,bcdr}, and
clustering \cite{my,ou,mwl}.  The inelastic nature of the collisions
remains crucially important in dilute settings and under vigorous
forcing where, in contrast with molecular gases, there is no energy
equipartition \cite{wp,fm} and the velocity distributions are
typically non-Maxwellian \cite{kwg,ep,ve,rm,ao,bbrtv}. In this letter,
we show that energy dissipation also results in self-sustaining
stationary states where energy cascades from large velocity scales to
small velocity scales, and we propose that these states may be
realized experimentally in driven granular gases.

Hydrodynamic theory of granular flows is formulated using
inelastic gases as a starting point \cite{pkh,jr,gzb,bdks,ig}.
Without forcing, dissipation is quantified via the energy balance
equation, $dT/dt=-\Gamma$, where $T=\langle v^2\rangle$ is the
temperature and $\Gamma$ the dissipation rate. Collisions
involving a pair of particles with relative velocity $\Delta v$
occur with rate $\propto(\Delta v)^\lambda$ and result in energy
loss $\propto(\Delta v)^2$. Assuming that a single velocity scale
characterizes the system, $\Gamma \sim \langle (\Delta
v)^{2+\lambda}\rangle \sim T^{1+\lambda/2}$, leads to Haff's
cooling law, $T\sim t^{-2/\lambda}$ (exponential decay occurs when
$\lambda=0$). Hence, the temperature decays indefinitely and the
velocity distribution $f(v)$ approaches the trivial steady-state
$f(v)\to \delta(v)$ where all particles are at rest, a stationary
state that can be considered as a dynamical fixed point. However,
the energy balance equation assumes a finite dissipation rate.
This need not be the case!

Our main result is that there is a general class of nontrivial
stationary states where the velocity distribution has an algebraic
high-energy tail
\begin{equation}
\label{powerlaw}
f(v)\sim v^{-\sigma}
\end{equation}
as $v\to \infty$. First, we obtain this result in one-dimension and
then, generalize it to arbitrary dimension.

Our system is an ensemble of identical particles that undergo
inelastic collisions
\begin{equation}
\label{law-1d}
v_{1,2}=pu_{1,2}+qu_{2,1}
\end{equation}
with $u_{1,2}$ ($v_{1,2}$) the pre-collision (post-collision)
velocities.  The collision parameters $p$ and $q$ obey $p+q=1$ so
momentum is conserved. In each collision, the relative velocity is
reduced by the restitution coefficient \hbox{$r=1-2p$} and the
energy loss equals \hbox{$pq(u_1-u_2)^2$}.  We consider general
collision rates \hbox{$|u_1-u_2|^\lambda$} with $\lambda$ the
homogeneity index. For particles interacting via the central
potential \hbox{$U(r)\sim r^{-a}$} then
\hbox{$\lambda=1-2\frac{d-1}{a}$} so \hbox{$\lambda\leq 1$}
\cite{rd}.  Hard-spheres, $\lambda=1$, model ordinary granular
media.  Maxwell molecules \cite{mhe}, \hbox{$\lambda=0$}, model
granular particles with dipole interactions, such as magnetic
particles or particles immersed in a fluid \cite{ia}.

We seek stationary velocity distributions $f(v)$ that obey the
Boltzmann equation
\begin{eqnarray}
\label{be} \iint du_1\,du_2\, |u_1-u_2|^\lambda\, f(u_1)f(u_2)
\qquad\qquad\qquad\\
\times\left[\delta(v-pu_1-qu_2)-\delta(v-u_1)\right]=0,\nonumber
\end{eqnarray}
reflecting balance of gain and loss due to collisions.

The case $\lambda=0$ is instructive as the velocity distribution can
be obtained explicitly. The Boltzmann equation is a simple convolution
because the collision rate is constant, and it is conveniently studied
using the Fourier transform \hbox{$F(k)=\int dv\, f(v)\,e^{ikv}$} that
satisfies the nonlinear, nonlocal equation \hbox{$F(k)=F(pk)F(qk)$}.
Normalization implies $F(0)=1$.  For elastic collisions ($p=0$) any
distribution is stationary, but this is just a one-dimensional
pathology. Generally, for any value of the restitution coefficient,
there is a family of solutions \hbox{$F(k)=\exp\big(-v_0|k|\big)$}
with arbitrary typical velocity $v_0$. As a result, the velocity
distribution is Lorentzian
\begin{equation}
\label{lorentz}
f(v)=\frac{1}{\pi v_0}\frac{1}{1+(v/v_0)^2}.
\end{equation}
We note that this distribution is independent of the restitution
coefficient.  The typical velocity $v_0$ is finite, but the
average energy is infinite.

This distribution does not evolve under the inelastic collision
dynamics and in particular, the energy density remains constant.  From
the energy balance equation, one might have expected that the velocity
distribution constantly narrows because particles dissipate energy,
but quite surprisingly, the heavy tail of the velocity distribution
acts as a heat bath, balancing the energy dissipation and maintaining
a steady velocity distribution.  We conclude that, in addition to the
trivial state where all particles are at rest, there is another fixed
point, a self-sustaining stationary state.

For general collision rules, the characteristic exponent $\sigma$ can
be found analytically.  For large $v$, the convolution in
Eq.~(\ref{be}) is governed by the product $f(u_1)f(u_2)$ with one of
the pre-collision velocities large and the other small because the
distribution decays sharply at large velocities. The Boltzmann
equation includes a loss term and a gain term.  For the loss term,
$u_1=v$ is large and $u_2$ is small. For the gain term, there are two
separate contributions. Either $v=pu_1$ and then $u_2$ is small or
$v=qu_2$ and then $u_1$ is small. In either case, the double integral
separates. The integral over the smaller velocity equals one.  With
the remaining integral over the larger velocity, denoted by $u$, the
nonlinear Boltzmann equation becomes linear for large $v$
\begin{eqnarray}
\label{linear-int-1d}
\int\! du\, |u|^\lambda f(u)
\left[\delta(v\!-\!pu)\!+\!\delta (v\!-\!qu)\!-\!\delta(v\!-\!u)\right]=0.\quad
\end{eqnarray}
Consequently, the tail of the distribution satisfies the functional
equation
\begin{eqnarray}
\label{linear-1d}
\frac{1}{p^{1+\lambda}}f\left(\frac{v}{p}\right)+
\frac{1}{q^{1+\lambda}}f\left(\frac{v}{q}\right)-f(v)=0
\end{eqnarray}
describing cascade of energy from large velocities to small ones,
$v\to (pv,qv)$.  The power-law decay (\ref{powerlaw}) satisfies this
equation when \hbox{$p^{\sigma-1-\lambda}+q^{\sigma-1-\lambda}=1$},
and since \hbox{$p+q=1$}, the characteristic exponent is
\begin{eqnarray}
\label{sigma-1d}
\sigma=2+\lambda.
\end{eqnarray}
In one-dimension, $\sigma$ is independent of the restitution
coefficient.

We comment that the full nonlinear Boltzmann equation assumes
molecular chaos as the two-particle velocity distribution is a product
of one-particle distributions. However, the equation governing the
tail is linear and it is valid under less stringent conditions. The
only requirement is that energetic particles are weakly correlated
with slower ones.

In arbitrary dimension $d$, the collision law is
\begin{equation}
\label{rule}
{\bf v}_1= {\bf u}_1-(1-p)
({\bf u}_1-{\bf u}_2)\!\cdot\!\hat{\bf n}\,\hat{\bf n}.
\end{equation}
Momentum transfer occurs only along the normal direction with
 $\hat{\bf n}$ a unit vector parallel to the impact direction
 connecting the particles. The energy dissipation equals
 \hbox{$p(1-p)|({\bf u}_1-{\bf u}_2)\!\cdot\!\hat{\bf n}|^2$} and the
 collision rate is \hbox{$|({\bf u}_1-{\bf u}_2)\!\cdot\!\hat{\bf
 n}|^\lambda$}.

We derive the linear equation governing the tail of the distribution
directly from the collision rule.  With the large velocity ${\bf u}$,
its magnitude \hbox{$u\equiv |{\bf u}|$} and \hbox{$\mu=(\hat{\bf
u}\cdot\hat{\bf n})^2$}, the collision rate is
\hbox{$u^\lambda\mu^{\lambda/2}$}. The cascade process is \hbox{$v\to
(\alpha v,\beta v)$} with the stretching parameters
\hbox{$\alpha=(1-p)\mu^{1/2}$} and \hbox{$\beta=[1-(1-p^2)\mu]^{1/2}$}
for \hbox{${\bf v}=(1-p){\bf u}\cdot\hat{\bf n}\,\hat{\bf n}$} and
\hbox{${\bf v}={\bf u}-(1-p){\bf u}\cdot\hat{\bf n}\,\hat{\bf n}$},
respectively \cite{beta}. Integrating over the impact angle $\hat{\bf
n}$ and the velocity magnitude $u$, Eq.~(\ref{linear-int-1d})
generalizes as follows
\begin{eqnarray*}
\left<\int\!du\, u^{\lambda+d-1}\!f(u)\!
\left[\delta(v\!-\!\alpha u)\!+\!\delta(v\!-\!\beta u)\!-\!f(v)\right]\!
\mu^{\lambda/2}\!\!\right>\!=\!0.
\end{eqnarray*}
Here, the brackets are used as shorthand for the angular
integration, \hbox{$\langle g\rangle =\int d\hat{\bf n}\,
g(\hat{\bf n})$}.  Integrating over $u$
yields
\begin{eqnarray}
\label{tail-eq}
\left<\left[\frac{1}{\alpha^{d+\lambda}}f\left(\frac{v}{\alpha}\right)+
\frac{1}{\beta^{d+\lambda}}f\left(\frac{v}{\beta}\right)-f(v)
\right]\!\mu^{\lambda/2}\!\right>=0.\quad
\end{eqnarray}
The power-law decay (\ref{powerlaw}) satisfies this equation
provided that the exponent $\sigma$ is root of the equation
$\left<\left(1-\alpha^{\sigma-d-\lambda}-
\beta^{\sigma-d-\lambda}\right)\mu^{\lambda/2}\right>=0$. The
angular integration is performed using spherical coordinates.
Given \hbox{$\mu=\cos^2\theta$}, then \hbox{$d\hat{\bf n}\propto
(\sin \theta)^{d-2}d\theta$}, and \hbox{$\langle \,g
\,\rangle\propto \int_0^1 d\mu\, \mu^{-{1\over
2}}(1-\mu)^{d-3\over 2}g(\mu)$}.  The equality for the exponent
$\sigma$ can be written in terms of the gamma function and the
hypergeometric function \cite{gr}
\begin{eqnarray}
\label{main}
\frac{1\!-\!{}_2F_1\!\left(\frac{d+\lambda-\sigma}{2},\frac{\lambda+1}{2},
\frac{d+\lambda}{2},1\!-\!p^2\right)}{(1-p)^{\sigma-d-\lambda}}\!=\!
\frac{\Gamma(\frac{\sigma-d+1}{2})\Gamma(\frac{d+\lambda}{2})}
{\Gamma(\frac{\sigma}{2})\Gamma(\frac{\lambda+1}{2})}.\quad
\end{eqnarray}

In the cascade process \hbox{$v\to(\alpha v,\beta v)$}, the total
velocity magnitude increases \hbox{($\alpha+\beta\geq 1$)} even though
the total energy decreases \hbox{($\alpha^2+\beta^2\leq 1$)}.
Utilizing these inequalities, we deduce the bounds \hbox{$1\leq
\sigma-d-\lambda\leq 2$} from the embedded equation for the
characteristic exponent. The lower bound (\ref{sigma-1d}) is realized
in one-dimension where collisions are head-on \hbox{($\mu=1$)}. The
upper bound is approached, \hbox{$\sigma\to d+2+\lambda$}, in the
quasi-elastic limit, $r\to 1$. This, combined with the known
Maxwellian distribution that occurs for elastic collisions,
demonstrates the singular nature of the quasi-elastic limit.
Interestingly, the energy may be either finite or infinite, depending
on whether $\sigma$ is larger or smaller than $d+2$. In either case,
the integral underlying the dissipation rate is divergent, a general
characteristic of the stationary distributions.

Generally, the exponent $\sigma$ increases monotonically with the
spatial dimension $d$, the homogeneity index $\lambda$, and the
restitution coefficient $r$.  For completely inelastic ($r=0$)
hard spheres, we find $\sigma=4.14922$ and $\sigma=5.23365$ for
$d=2$, $3$, respectively. The exponents for completely inelastic
Maxwell molecules are $\sigma=3.19520$ and $\sigma=4.28807$ for
$d=2$, $3$. These values provide lower bounds on $\sigma$ with
respect to $r$, as shown in figure 1.

\begin{figure}[ht]
\includegraphics*[width=0.44\textwidth]{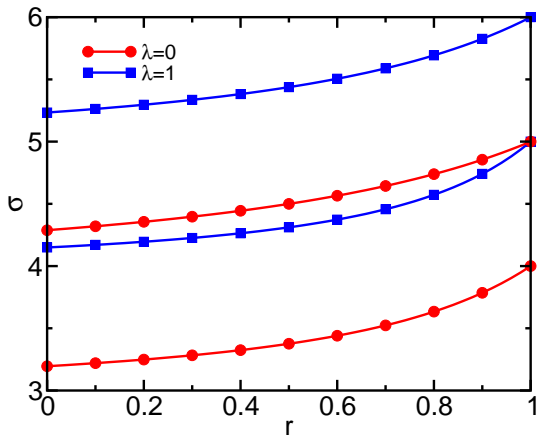}
\caption{The exponent $\sigma$ as a function of the restitution
coefficient $r$ for $\lambda=0$ (circles) and $\lambda=1$
(squares). The lower curves correspond to $d=2$ and the upper to
$d=3$.}
\end{figure}

\begin{figure}[ht]
\includegraphics*[width=0.44\textwidth]{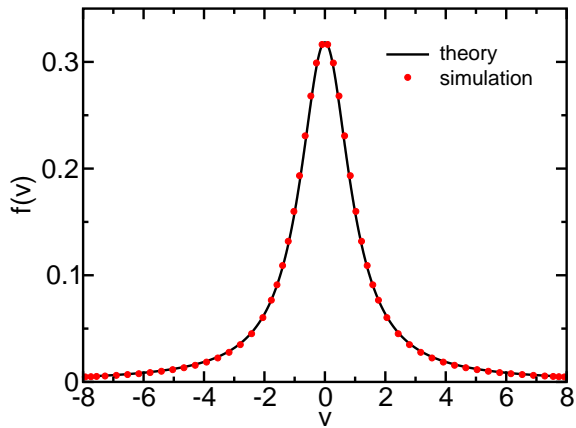}
\caption{The velocity distribution $f(v)$ versus $v$ for Maxwell
molecules in one-dimension.  The velocity was rescaled to compare with
an ordinary Lorentzian.}
\end{figure}

A remarkable feature is that the characteristic exponent changes
continuously with the parameters $d$, $\lambda$, and $r$. Hence, the
tails are not universal. The power-law decay stands in contrast with
the stretched exponentials, \hbox{$f(v)\sim \exp(-v^{\nu})$}, with
$\nu=\lambda$ and $\nu=1+\lambda/2$, respectively, for unforced and
thermally forced inelastic gases \cite{ep,ve,rm,ao}. Previously,
algebraic tails with different exponents were found but for
non-stationary distributions describing freely cooling Maxwell
molecules \cite{bk,bmp,eb}.

\begin{figure}[ht]
\includegraphics*[width=0.43\textwidth]{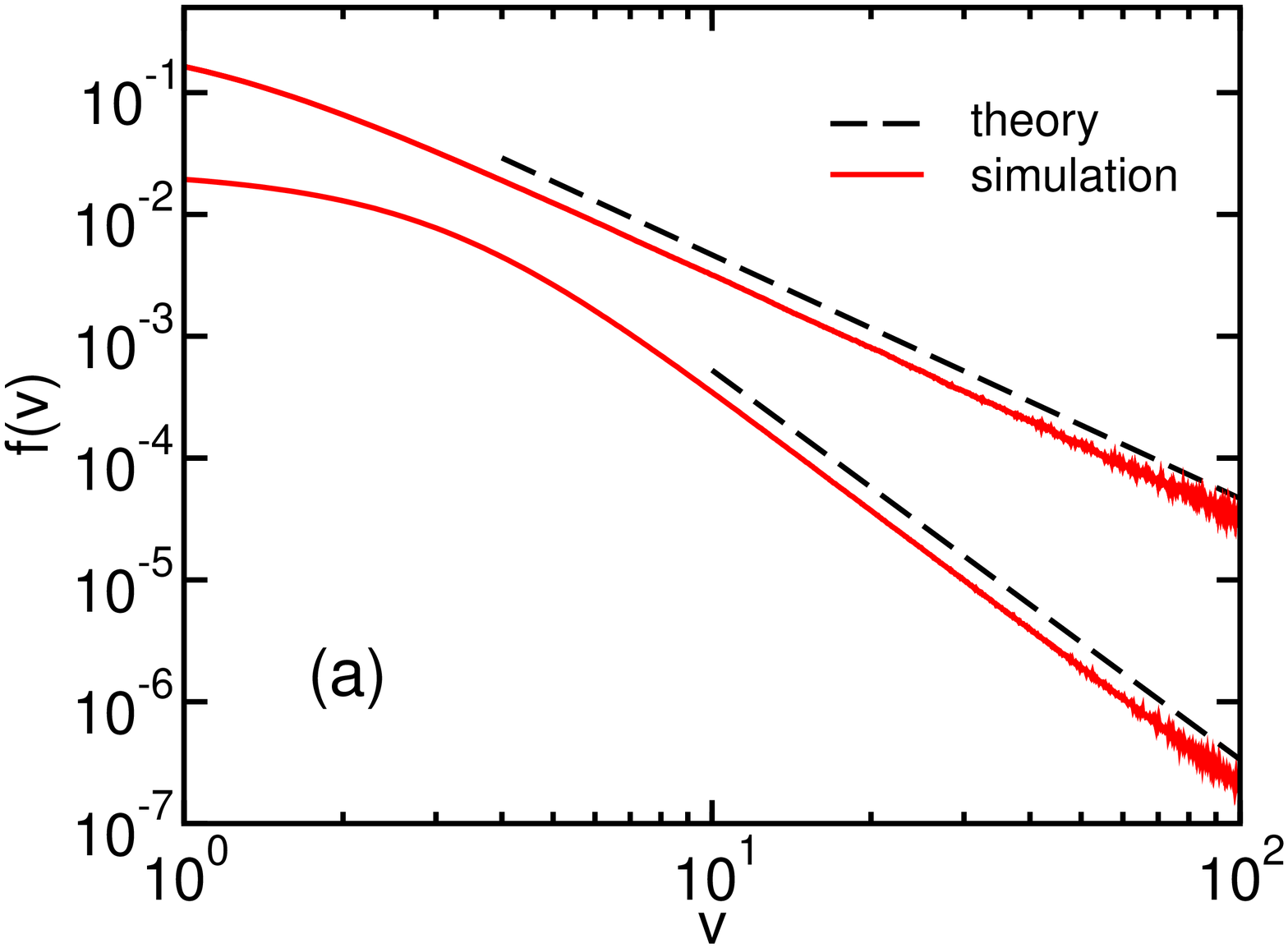}
\includegraphics*[width=0.43\textwidth]{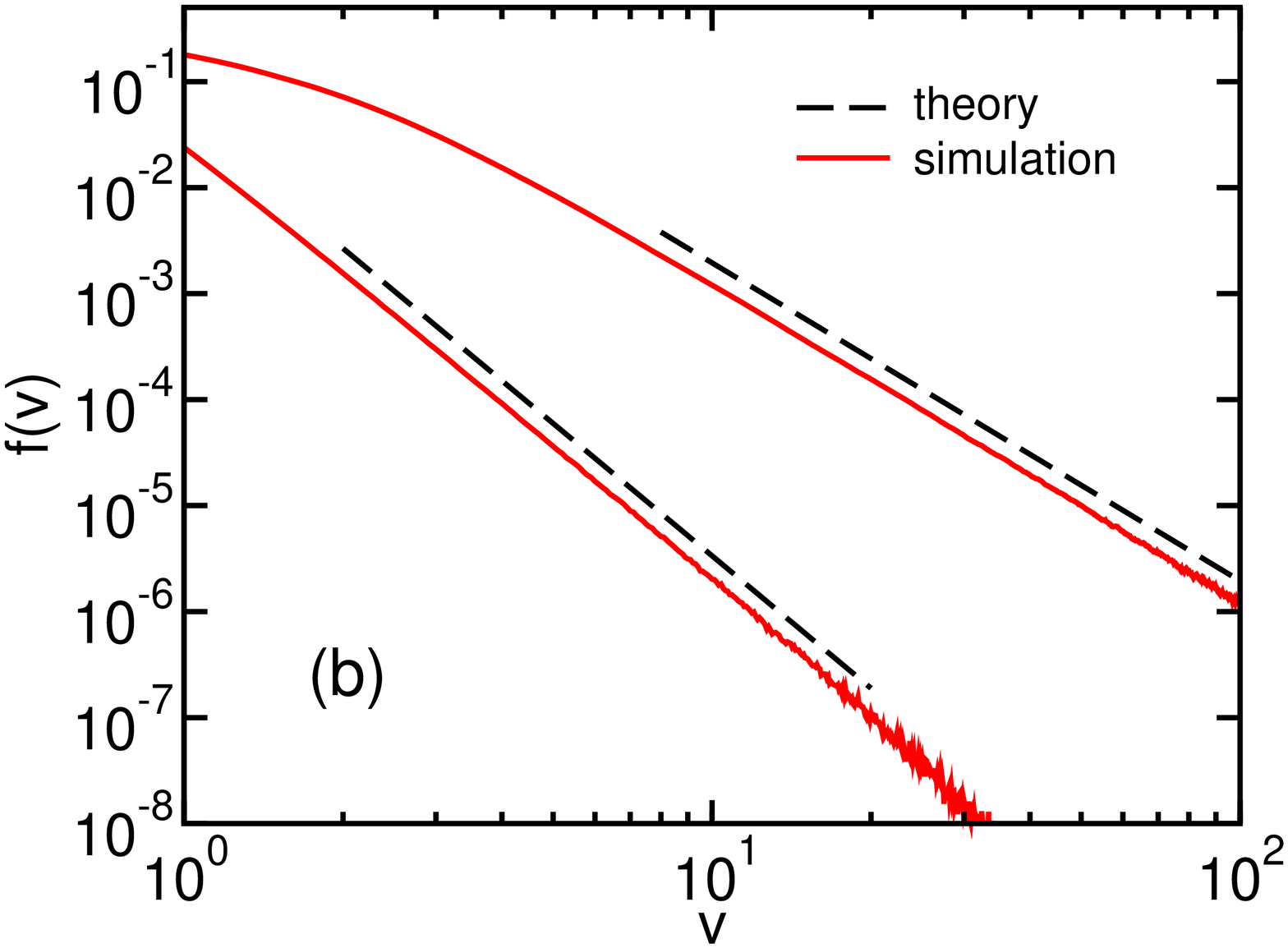}
\caption{The velocity distribution $f(v)$ versus $v$ for Maxwell
molecules (a) and hard spheres (b) in one-dimension (upper lines) and
two-dimension (lower lines). The theoretical prediction is shown for
reference.}
\end{figure}

These steady-state distributions can be realized, up to some cutoff,
in finite systems of driven inelastic particles. The key is to inject
energy at a very large velocity scale. Numerically, we used the
following simulation. Initially, we start with an innocuous velocity
distribution, uniform with support in $[-1:1]$. Particles collide
inelastically, and an energy loss ``counter'' keeps track of the
aggregate energy loss.  With rate $\gamma$, small compared with the
collision rate, a particle is selected at random and it is energized
by an amount equal to the aggregate energy loss. Subsequently, the
energy loss counter is reset to zero. The rationale behind this
simulation is that the total energy remains practically constant and
more importantly, that energy is injected only at the tail of the
distribution. In effect, this procedure does not alter the purely
dissipative dynamics, except for setting a scale for the most
energetic particles.

We simulated completely inelastic hard-spheres and Maxwell molecules
in one- and two-dimensions.  We used $N=10^7$ ($N=10^5$) particles and
injection rate $\gamma=10^{-4}$ ($\gamma=10^{-2}$) for $\lambda=0$
($\lambda=1$). We verified that: (i) after a transient, the velocity
distribution becomes stationary, (ii) the velocity distribution is
Lorentzian for one-dimensional Maxwell molecules (figure 2), (iii) the
velocity distribution has an algebraic tail, and (iv) $\sigma$ is
consistent with theory (figure 3).

The very same stationary distributions can be reached using other
simulation methods.  For example, particles may be re-energized
such that their velocity is drawn from a Maxwellian distribution
with a typical energy proportional to the system size (the data
presented for one-dimensional hard-spheres are from such a
simulation). We conclude that the simulations confirm the
existence of the nontrivial stationary states with power-law
tails. These steady-states are stable fixed points as the system
is driven into them even when starting from compact distributions.
Moreover, stability analysis of the time dependent version of
Eq.~(\ref{tail-eq}) shows that the stationary distribution
(\ref{powerlaw}) is stable with respect to perturbations
consisting of steeper algebraic tails.

Clearly, if $f(v)$ is a steady-state so is $v_0^{-d}f(v/v_0)$ for an
arbitrary typical velocity $v_0$. The injection protocol selects
$v_0$.  Suppose that particles are boosted at rate $\gamma$ per
particle to velocity $V$, a scale that sets an upper cutoff on the
velocity distribution \cite{finite}. The energy injection rate is
$\gamma V^2$, and the energy dissipation rate is $\Gamma\sim \langle
v^{2+\lambda}\rangle$ with \hbox{$\langle g\rangle \equiv \int^V dv
\,v^{d-1} f(v) g(v)$}. Energy balance, $\Gamma\sim \gamma V^2$,
relates the injection rate, the injection scale, and the typical
velocity scale, $\gamma\sim V^\lambda (V/v_0)^{d-\sigma}$.

In our simulations, energy is maintained constant, $\langle
v^2\rangle\sim 1$. When $\sigma<d+2$, the constant energy constraint
implies $V^{d+2-\sigma}\sim v_0^{d-\sigma}$, that combined with energy
balance reveals how the maximal velocity, $V\sim
\gamma^{-1/(2-\lambda)}$, and the typical velocity, $v_0\sim
\gamma^{\frac{d+2-\sigma}{(\sigma-d)(2-\lambda)}}$, scale with the
injection rate. When $\sigma>d+2$, the initial conditions set the
typical velocity: $\langle v^2\rangle\sim v_0^2$ implies $v_0\sim 1$,
and energy balance yields $V\sim \gamma^{-1/(\sigma-d-\lambda)}$. The
simulations are consistent with these scaling laws. For example,
$V\approx 10^2$ and $v_0\approx 10^{-2}$ for one-dimensional
Maxwell-molecule simulations with $\gamma=10^{-4}$. This, combined
with the simulations, demonstrates that finite energy and infinite
energy cases differ quantitatively but not qualitatively.

In conclusion, we have shown that inelastic gases have nontrivial
stationary states where the velocity distribution has an algebraic
high-energy tail.  These steady-states are generic in that they
exist for arbitrary collision rates, arbitrary collision rules,
and arbitrary spatial dimensions. Typically, energy is injected at
large velocity scales, and it is dissipated over a range of
smaller velocities. Qualitatively, this energy cascade picture is
analogous to fluid turbulence. However, without energy
conservation, there is an important difference with the
conventional Kolmogorov spectra: the characteristic exponents are
not universal and they do not follow from dimensional analysis
\cite{zlf}.

In an infinite system, there is perfect balance between collisional
loss and gain and the high-energy tail enables the stationary state,
while in finite systems, energy injection maintains the
steady-state. The latter is relevant to granular gas experiments that
are typically performed in steady-state conditions. We propose that
stable distributions and energy cascades may be realized using an
experimental set-up in which energy is added at large velocity scales
via rare but powerful energy injections.

We thank A.~Baldassarri, N.~Menon, and H.~A.~Rose for useful
discussions.  We acknowledge DOE W-7405-ENG-36 and NSF DMR-0242402
for support of this work.


\begin{thebibliography}{99}

 \bibitem{gz}
      I.~Goldhirsch, and G.~Zanetti,
      Phys. Rev. Lett. {\bf 70}, 1619 (1993).

\bibitem{km}
      E.~Khain and B.~Meerson,
      Europhys. Lett. {\bf 65}, 193 (2004).

\bibitem{rbss}
      E.~C.~Rericha, C.~Bizon, M.~D.~Shattuck, and H.~L.~Swinney,
      Phys.\ Rev.\ Lett. {\bf 88}, 014302 (2002).

\bibitem{slk}
      A.~Samadani, L.~Mahadevan, and A.~Kudrolli,
      J. Fluid Mech. {\bf 452}, 293 (2002).

\bibitem{bcdr}
      E.~Ben-Naim, S.~Y.~Chen, G.~D.~Doolen, and S.~Redner,
      Phys.\  Rev.\  Lett.\ {\bf 83}, 4069 (1999).

\bibitem{my}
      S.~McNamara and W.~R. Young, Phys Fluids A {\bf 4}, 496 (1992).

\bibitem{ou}
      J.~S.~Olafsen and J.~S.~Urbach,
      Phys.\ Rev.\ Lett. {\bf 81}, 4369 (1998).

\bibitem{mwl}
      D.~van der Meer, K.~van der Weele, and D.~Lohse,
      Phys. Rev. Lett {\bf 88}, 174302 (2002).

\bibitem{wp}
      R. D. Wildman and D. J. Parker,
      Phys.\ Rev.\ Lett. {\bf 88}, 064301 (2002).

\bibitem{fm}
      K.~Feitosa and N.~Menon,
      Phys. Rev. Lett. {\bf 88}, 198301 (2002).

\bibitem{kwg}
      A.~Kudrolli, M.~Wolpert, and J.~P.~Gollub,
      Phys.\ Rev.\ Lett. {\bf 78}, 1383 (1997).

\bibitem{ep}
      S.~E.~Esipov and T.~P\"oschel,
      J. Stat.\ Phys. {\bf 86}, 1385 (1997).

\bibitem{ve}
      T.~P.~C.~van~ Noije and M.~H.~Ernst,
      Gran. Matt. {\bf 1}, 57 (1998).

\bibitem{rm}
      F.~Rouyer and N.~Menon,
      Phys.\ Rev.\ Lett. {\bf 85}, 3676 (2000)

\bibitem{ao}
      I.~S.~Aranson and J.~S.~Olafsen
      Phys. Rev. E {\bf 66}, 061302 (2002).

\bibitem{bbrtv}
      A.~Barrat, T.~Biben, Z.~R\'acz, E.~Trizac,
      and F.~van Wijland, J. Phys.\ A {\bf 35}, 463 (2002).

\bibitem{pkh}
      P.~K. Haff, J. Fluid Mech. {\bf 134}, 401 (1983).

\bibitem{jr}
      J.~T.~Jenkins and M.~W.~Richman,
      Phys. Fluids {\bf 28}, 3485 (1985).

\bibitem{gzb}
       E.~L.~Grossman, T.~Zhou, and E.~Ben-Naim,
       Phys.\ Rev.\ E {\bf 55}, 4200 (1997).

\bibitem{bdks}
     J.~J.~Brey, J.~W.~Dufty, C.~S.~Kim, and A.~Santos,
     Phys. Rev. E {\bf 58}, 4638 (1998).

\bibitem{ig}
     I.~Goldhirsch,
     Ann. Rev. Fluid. Mech. {\bf 35}, 267 (2003).

\bibitem{rd}
      P.~R\'esibois and M.~de~Leener, {\it Classical Kinetic
      Theory of Fluids} (John Wiley, New York, 1977).

\bibitem{mhe}
      M.~H.~Ernst,
      Phys. Reports {\bf 78}, 1 (1981).

\bibitem{ia}
     I.~S.~Aranson, private communication. 

\bibitem{beta} 
     The parameter $\beta$ is obtained by writing ${\bf
     w}={\bf v}-{\bf u}$ and employing the collision rule, $w=-{\bf
     w}\cdot \hat{\bf n}=(1-p){\bf u}\cdot \hat{\bf
     n}=(1-p)u\mu^{1/2}$ and then using $v^2=u^2+w^2-2uw\mu^{1/2}$.

\bibitem{gr}
     I.~S.~Gradshteyn and I.~M.~Ryzhik,
     {\it Table of Integrals, Series, and Products},
     (Academic Press, New York, 1972).

\bibitem{bk}
      E.~Ben-Naim and P.~L.~Krapivsky,
      Phys. Rev. E {\bf 61}, R5 (2000);
      P.~L.~Krapivsky and E.~Ben-Naim,
      J.\ Phys.\ A {\bf 35}, L147 (2002).

\bibitem{bmp}
      A.~Baldassarri, U.~M.~B.~Marconi, and A.~Puglisi,
      Europhys.\ Lett. {\bf 58}, 14 (2002).

\bibitem{eb}
      M.~H.~Ernst and R.~Brito,
      Europhys. Lett. {\bf 58}, 182 (2002).

\bibitem{finite}
      The system size must be large enough so that finite size scaling
      effects play no role.

\bibitem{zlf}
      V.~Zakharov, V.~Lvov, and G.~Falkovich,
      {\it Kolmogorv Spectra of Turbulence}
      (Springer-Verlag, Berlin, 1992).

\end{thebibliography}
\end{document}